# Widening Perspectives: The Intellectual and Social Benefits of Astrobiology, Big History, and the Exploration of Space

Ian A. Crawford

*Department of Earth and Planetary Sciences,*
*Birkbeck College, University of London, UK*

## Abstract

Astrobiology is the field of science devoted to searching for life elsewhere in the Universe. It is inherently inter-disciplinary, integrating results from multiple fields of science, and in this respect has strong synergies with 'big history'. I argue that big history and astrobiology are both acting to widen human perspectives in intellectually and socially beneficial directions, especially by enhancing public awareness of cosmic and evolutionary world-views. I will further argue that these perspectives have important implications for the social and political organ-isation of humanity, including the eventual political unification of our planet. Astrobiology and big history are also concerned with the future of humanity, and I will argue that this future will be culturally and intellectually enriched if it includes the exploration of the universe around us.

*I*t is only when the different scientific disciplines and the different specialities choose to interact, and only when all cultures and states recognize that they have common interests, that humanity can evolve towards one single co-operative society. (Aerts et al., 1994; p. 20)

## Introduction

Astrobiology and 'big history' are two relatively new intellectual disciplines, the former focussed on searching for life elsewhere in the universe and the latter on integrating human history with the wider history of the cosmos. Despite some differences in emphasis these two disciplines share much in common, not least their interdisciplinarity and the cosmic and evolutionary perspectives that they both engender. In this essay I will explore the relationships between astrobiology and big history and argue that both are acting to wid-en human perspectives in intellectually and socially beneficial directions. These include stimulating the (partial) re-integration of scientific disciplines after a period of extreme specialisation, and the (again partial) breaking down of barriers that exist between the sciences and the humanities. In addition, both disci-plines act to enhance public awareness of cosmic and evolutionary perspectives which, I will argue, consti-tute a strong, if implicit, argument for the eventual po-litical unification of humanity. Astrobiology and big history are also concerned with the *future* of humanity, and I will make the case that the future will be cultur-ally and intellectually richer if it includes an ambitious programme of space exploration. Not only will the ex-ploration of space further reinforce socially beneficial cosmic perspectives, but ultimately it may be the only way for human (and post-human) societies to avoid the intellectual stagnation once predicted for the 'End of History'.

**Astrobiology and Big History**

The International Big History Association adopts the following working definition for the discipline:

> Big History seeks to understand the integrated history of the cosmos, Earth, life and humanity, using the best available evidence and scholarly methods.[1]

This is strikingly similar to a common working definition of the comparably recent discipline of astrobiology:

> The scientific study of the possible origin, distribution, evolution, and future of life in the universe, including that on Earth, using a combination of methods from biology, chemistry, and astronomy.[2]

Although the term "astrobiology" dates from 1953 (see, e.g., Cockell, 2001), it is only in the last 25 years or so that it has become firmly established as a scientific discipline, with the appearance of dedicated textbooks, journals, and university courses. The field is inherently interdisciplinary because any serious attempt to understand the prevalence and distribution of life in the universe requires familiarity with, at least, the established scientific disciplines of astronomy, biology, chemistry and geology (as well as established interdisciplinary combinations among these sciences, e.g., astrophysics, biochemistry, evolutionary biology, geochemistry, palaeontology, and planetary science). In order to illustrate the interdisciplinary nature of astrobiology more clearly, Table 1 summarises the syllabus of the undergraduate module "Introduction to Astrobiology" that the author has taught at Birkbeck College, University of London, since 2004.[3]

| Table 1 | | |
|---|---|---|
| Week | Topic | Most relevant scientific field(s) |
| 1 | Origin and distribution of the chemical elements | Astronomy/Astrophysics |
| 2 | Conditions in the early Solar System | Astronomy, Planetary science |
| 3 | Earliest evidence for life on Earth | Geology, Palaeontology |
| 4 | Biological basics | Biology, Biochemistry |
| 5 | Pre-biological chemical evolution/Origin of life | Geochemistry, Biology, Biochemistry |
| 6 | History of life on Earth | Palaeontology/Evolutionary biology |
| 7 | Requirements for life | Biology/Biochemistry/Geochemistry |
| 8 | Prospects for life on Mars | Planetary science/Geochemistry/Biology |
| 9 | Life elsewhere in the Solar System | Planetary science/Geochemistry/Biology |
| 10 | Detection and habitability of exoplanets | Astronomy/Planetary science |
| 11 | Search for extraterrestrial intelligence | Astronomy |

**Table 1**: Syllabus of the Birkbeck College "Introduction to Astrobiology" module (each week comprises three hours of face-to-face teaching).

---

1   https://bighistory.org/ (accessed 18 November, 2018); see also Rodrigue (2017).

2   https://www.thefreedictionary.com/astrobiology (accessed 18 November, 2018).

3   http://www.bbk.ac.uk/study/modules/easc/EASC064H5 (accessed 18 November, 2018).

A glance at Table 1 indicates that approximately half of this undergraduate astrobiology module could equally be described as big history[4]. With the exception of the material covered in Week 4, which is included to ensure that non-biology students are familiar with at least the basics of biological knowledge, the material covered in Weeks 1-6 is all essentially 'historical' in nature (albeit invoking a range of scientific disciplines) and is invariably covered in the first few chapters of standard big history texts (e.g. Christian, 2004, 2018; Brown, 2007; Christian et al., 2014; Spier, 2015). After this point astrobiology necessarily diverges from big history, with the former branching out to look for life elsewhere in the Universe while the latter continues the historical narrative to include the evolution of *Homo sapiens*, human societies and human culture.

The links between astrobiology and big history may be further illustrated by means of a personal anecdote: the first half of the astrobiology syllabus outlined in Table 1 is based on an earlier course entitled "Cosmic Perspectives for World History" that I devised for the City University's extramural programme in 1994 (see Figure 1). At the time I was unaware of big history as such, although Christian (1991) had already coined the term. I was, however, partly inspired by G.S. Kutter's (1986) book *The Universe and Life*, which is often identified as a big history precursor (Rodrigue, 2017). In retrospect, it is clear that this early 'Cosmic Perspectives' course, which in time led to the Birkbeck College undergraduate module in astrobiology, was big history in all but name. This anecdote reinforces observations already made by others that the early years of big history were characterised by individuals and small groups working independently. It seems that by the late 20th century big history was an idea whose 'time had come', although of course the subject has much deeper roots (see, e.g., Spier, 2015; Rodrigue, 2017; Katerberg, 2018).[5]

William Katerberg (2018) has recently argued that the academic fields closest to big history are deep history (where 'deep' here refers to human pre-history), evolutionary history, and ecological economics. Based on the discussion above, however, I suggest that astrobiology is an even closer match, both in terms of content and perspective (where there is considerable overlap), but also in the way both disciplines have struggled, eventually successfully, for academic recognition over the last quarter of a century.

Much more important than the origins of interdisciplinary subjects like astrobiology and big history, however, is the extent to which they can have lasting intellectual and societal benefits. Because the academic and intellectual benefits of these subjects, and what I perceive as their wider societal benefits, are rather different (albeit interconnected) they will be addressed separately below.

## Intellectual Benefits of Big History and Astrobiology[6]

The main academic and intellectual benefits of both astrobiology and big history (and related disciplines) arise from their inherent interdisciplinarity. In the case of astrobiology these benefits have already been noted by several authors (e.g., Connell et al., 2000; Race et al., 2012), and mostly result from interactions between scientific disciplines. For example, astrobiology forces astronomers to work with biologists and geologists in the pursuit of finding life elsewhere in the universe. By producing broadly knowledgeable scientists, familiar with multiple aspects of the natural world, astrobiology stimulating intellectual activity outside the normal scope of the physical sciences, including theoretical work in anthropology, ethics, linguistics, philosophy, and theology (e.g., Bertka, 2009; Dick & Lupisella, 2009; Race et al., 2012; Dunér et al., 2013; Impey et al., 2013; Vakoch, 2013, 2014; Dick, 2018). To this extent, astrobiology is well-placed, if only partially, to help heal the rift between science and the humanities identified sixty years ago by C.P. Snow in his famous 1959 Rede Lecture at the University of Cambridge (Snow, 1963; pp. 1-51).

---

4    See also Dick (2018), pp. 169, 235, 311.

5    If I may be permitted an additional personal anecdote: having had the proposal for a course on 'Cosmic Perspectives' accepted by the City University in 1993, I started writing it while working at the Anglo-Australian Observatory, then based in Epping, a northern suburb of Sydney. This was (almost literally!) a stone's throw from Macquarie University, where David Christian was already developing his big history perspective, although neither of us knew of each other's existence.

6    The astrobiology side of the discussion in the following two sections draws on an earlier publication (Crawford, 2018a).

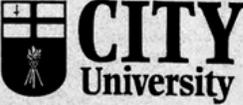

**Figure 1.** The syllabus of a course on "Cosmic Perspectives for World History" taught by the author at the City University, London, in the academic years 1994-95 and 1995-96. Image by the author.

Similar arguments have been advanced for big history, although there are some differences in emphasis (e.g. Christian, this volume). Big history clearly has the potential to stimulate research activity in the natural sciences, on which it relies for much of its historical narrative, but in origin, and perhaps especially in outlook, big history is closer to the humanities than interdisciplinary natural sciences such as astrobiology. To my mind, this strengthens the synergies between them, not least because it means that big history is even better placed to bridge Snow's "two cultures" divide.

The synergies between big history and astrobiology are perhaps most apparent when it comes to interdisciplinary education, and this may indeed prove to be one of the most important legacies of both disciplines. Snow himself explicitly recognized the importance of interdisciplinary education when he returned to the problem of the "two cultures" with *Two Cultures: A Second Look* (Snow, 1963; p. 61):

is therefore helping to re-unify the sciences after a long period of intense specialization. Moreover, by considering questions related to the philosophical and cultural implications of the discovery (or non-discovery) of extraterrestrial life, astrobiology is also In the conditions of our age, or any age which we can foresee, Renaissance man is not possible. But we can do something. The chief means open to us is education. … There is no excuse for letting another generation be as vastly ignorant, or as devoid of understanding and sympathy, as we are ourselves.

Interestingly, in the same year as Snow's *Second Look* appeared, the astronomer Harlow Shapley also made a powerful plea for interdisciplinary education. Shapley went as far as to characterise the 'vertical' separation of academic disciplines as "education-defeating" (Shapley, 1963; p. 134) and proposed that an ideal undergraduate historical curriculum would present the history of the universe and mankind as deduced from geology, cosmogony, paleontology, anthropology, comparative neurology, political history, and so on. … wide integration is the essential key (Shapley, 1963; pp. 135-6).

In 2009, the art historian Martin Kemp contributed an article in the scientific journal *Nature* to mark the 50th anniversary of Snow's original lecture. He concluded that the main problem was not so much a division between "two monolithic 'cultures' of science and humanities", but the "narrow specialisation of all disciplines." As he put it (Kemp, 2009):

It is the perceived need for intense specialization of any kind – in history or physics, in languages or biology – that needs to be tackled. …. What is needed is an education that inculcates a broad mutual understanding of the nature of the various fields of research.

This line of thinking has been taken up by others. For example, in an article stressing the desirability of producing scientifically minded citizens, Erika Offerdahl (2013) observed:

The structure of undergraduate curricula and courses tends to compartmentalize science into discrete disciplines that focus on particular questions rather than an integrated, interdisciplinary way of understanding the world, let alone any discussion of the societal implications of the science.

If nothing else, big history (and related interdisciplinary subjects such as astrobiology) can provide *exactly* this kind of interdisciplinary education, and do so in a manner that students of all ages find very engaging (e.g., Chaisson, 2014; Katerberg, 2018; Voros, 2018; Bohan, this volume). As Snow (1963; p. 61) himself noted, this will necessitate revising school and university curricula around the world, but the benefits of doing so are likely to be considerable (e.g., Katerberg, 2018; Bohan, this volume; Christian, this volume).

**Expanding Worldviews**

Transcending the academic, intellectual, and even practical benefits of a broadly-educated citizenry, the *perspectives* provided by astrobiology and big history may result in positive influences over a wide range of societal and political concerns. In an earlier article (Crawford, 2018a), I argued that wider public engagement with, and knowledge of, the topics covered by astrobiology (Table 1) would lead to beneficial social and political consequences. Based on the discussion above, it seems clear that these arguments are even stronger in the case of big history, which covers much of the same ground while explicitly articulating an evolutionary perspective rooted in deep time.

The key point relates to the broadening and deepening of worldviews resulting from increased public awareness of cosmic and evolutionary perspectives. Here, I adopt the definition of a worldview given by Diederik Aerts and colleagues in their excellent and important monograph on *World*

*Views: From Fragmentation to Integration* (Aerts et al., 1994; p. 9):

> A world view is a system of co-ordinates or a frame of reference in which everything presented to us by our diverse experiences can be placed. It is a symbolic system of representation that allows us to integrate everything we know about the world and ourselves into a global picture, one that illuminates reality as it is presented to us within a certain culture.

Aerts et al. (p. 8) also note that:

> World views …. have a strongly motivating and inspiring function. A socially shared view of the whole gives a culture a sense of direction, confidence and self-esteem.

Unfortunately, at present, and in some quarters increasingly, the worldviews of many people are dominated by narrow nationalistic and religious ideologies. Although historically some of these restrictive, and often mutually exclusive, worldviews may have had (local) societal benefits, and a propensity to hold them may have evolved naturally through group selection in humanity's distant past (e.g. Wallace, 1871, p.313; Darwin, 1874, p. 64; Wilson, 2012), they are potentially disastrous at a time of growing global interdependence. Our world faces many global problems (including, but not limited to, proliferation of weapons of mass destruction, climate change, pollution, loss of biodiversity, over exploitation of the 'global commons', and insufficient provision of food, water and sanitation for millions of people) that can only be satisfactorily addressed through concerted global action. However, meaningful global action will be, and is being, impeded by nationalistic and other essentially tribal worldviews, in which a sense of global identity and responsibility is lacking (or even denied). As Aerts et al. (p. 5) put it:

> It is our conviction that the time has come to make a conscious effort towards the construction of global world views, in order to overcome this situation of fragmentation. … It is precisely because we lack such global views of the world that our ability even to start looking for lasting solutions to these problems is limited.

There is therefore a pressing need to find unifying cosmopolitan perspectives that can counter the divisive and exclusionary worldviews of the past. In identifying such unifying worldviews, it will be essential that they are based on factual foundations that everyone can accept, and this is where big history and related disciplines are well-placed to help.

Spier (2016) has argued that big history should not be taken as an all-embracing worldview from which ethical implications can legitimately be drawn. He is undoubtedly correct that normative considerations cannot logically be derived from a factual history of the Universe such as big history seeks to provide. However, this does not mean that big history cannot provide a worldview (or, at least, part of a worldview) in the sense developed by Aerts et al. (1994), and that this worldview, once grasped, will not influence human behaviour. Indeed, the recognition that fact-based universal histories have ethical, and even political, implications has long been a significant motivation for constructing them. For example, in 1844 Robert Chambers published (anonymously) his *Vestiges of the Natural History of Creation*, which is perhaps the first serious attempt to create a (pre-Darwinian) evolutionary history of the Universe and humanity's place within it. Chambers himself certainly saw it as such, writing (p. 388):

> As far as I am aware [my book] is the first attempt to connect the natural sciences to a history of creation…. My sincere desire … was to give the true view of the history of nature.

*Vestiges* caused a huge sensation at the time (Secord, 2000), and the following year Chambers felt the need to offer some 'Explanations' (Chambers, 1845). In the course of this (p. 184) he explicitly drew the ethical implication that the "new view of nature" articulated in *Vestiges* could contribute to:

> Establishing the universal brotherhood and social communion of man. And not only this, but it extends the principle of humanity to the other meaner creatures also. Life is everywhere ONE.[7]

This quotation is especially significant because it shows that Chambers was concerned not just with laying a foundation for "the universal brotherhood and social communion of man", but also his expectation that a proper understanding of cosmic and evolutionary perspectives would have ethical implications for relations with other living things (and to this extent anticipates Peter Singer's (1981) concept of an 'expanding circle' of ethical progress).

The year following the publication of *Vestiges*, Alexander von Humboldt (1845) published his first volume of *Cosmos*, which also combined many different aspects of knowledge into an integrated view of humanity's place in the universe (albeit without the evolutionary emphasis of *Vestiges*). Humboldt's perspective was also seen to have unifying societal implications by at least some contemporaries, with the American physician and author James Whelpley (1846) noting that "the individual is made to feel that he is connected, by the very nature and substance of his body, with every part of the universe", and drawing the implication (p. 603) that:

> If the world is ever to be harmonized it must be through a community of knowledge, for there is no other universal or non-exclusive principle in the nature of man.

It appears that Whelpley had a sense that humanity might be able to "harmonize" itself socially and politically if it could only agree on a common integrated worldview of the kind Humboldt had developed. Several 20th Century advocates for what we might today call a 'big historical' worldview have likewise drawn attention to the societal benefits of the resulting cosmopolitan perspectives. H.G. Wells' *The Outline of History*, written in the appalling aftermath of the First World War, is arguably the foremost example, and Wells (1920, p. v) left no doubt about his reasons for writing it:

> The need for a common knowledge of the general facts of human history throughout the world has become very evident during the tragic happenings of the last few years …. There can be no common peace and prosperity without common historical ideas. Without such ideas to hold them together in harmonious co-operation, with nothing but narrow, selfish, and conflicting nationalist traditions, races and peoples are bound to drift towards conflict and destruction.

These considerations famously led Wells to conclude (p. 608) that "human history becomes more and more a race between education and catastrophe." He was convinced that every thinking person should do what they can to help win this race, and that finding a common historical perspective was the key (p. 603):

> The essential task of men of goodwill in all states and countries remains the same, it is an educational task, and its very essence is to bring to the minds of all men everywhere, as a necessary basis for world cooperation, *a new telling and interpretation, a common interpretation of history*.[8]

Other examples of arguments for the societal benefits of big historical/astrobiological perspectives include works by the astronomers Harlow Shapley and Hubert Reeves. Shapley, in particular, dedicated much of his career to popularising the cultural benefits of a cosmic perspective (see Palmeri, 2009) and began the preface of his book *The View from a Distant Star* (Shapley, 1963; p. 5) by noting:

---

7   Capitals in the original.

8   Emphasis in the original.

Mankind is made of star stuff, ruled by universal laws. The thread of cosmic evolution runs through his history. [9]

Shapley argued that this vast perspective could, indeed *should*, "incite orientating thoughts" (see pp. 38, 93, 161) that would, among other benefits, help "take us through the present and future predicaments" (p. 97) facing humanity. In his book *The Hour of Our Delight: Cosmic Evolution, Order and Complexity*, Reeves (1991) was similarly motivated by potential societal benefits arising from a knowledge of cosmic evolution and by the hope that the resulting "sense of wonder" would help turn humanity away from violence, conflict, and, especially, nuclear war. Reflecting on the contrast between the wonder of cosmic evolution revealed by modern science, and the often absurd pointlessness of human conflict, he wrote "The awakening of a sense of wonder and delight is the best antidote to absurdity at all levels" (Reeves, 1991; p. 8), and went on to propose that an understanding of cosmic evolution evokes an argument for human solidarity and dignity (p. 185):

A new vision of humanity emerges from contemporary scientific knowledge. Though mankind can no longer pretend to be the center of the world, our new position gives us our real dignity. … we occupy the top level of the pyramid of nature's organised entities. We reached this level after a gestation period of fifteen billion years, in which all of the cosmic phenomena participated. *All human beings, regardless of their origin, have an equal claim to this dignity*. The respect for human rights implies also an awareness of the importance of every individual in the history of the universe. [10]

Perhaps the clearest recent enunciation of why the perspectives provided by big history and related disciplines have the potential to help unite humanity was made by the biologist Ursula Goodenough in her 1998 book *The Sacred Depths of Nature* (p. xvi):

Any global tradition needs to begin with a shared worldview: a culture-independent, globally accepted consensus as to how things are. … our scientific account of nature, an account that can be called The Epic of Evolution. … this is the story, the one story, that has the potential to unite us, because it happens to be true.[11]

Given the potential importance of developing such a unified worldview, it would be desirable to assess empirically the extent to which the teaching of 'the epic of evolution' (which is essentially big history by another name) can achieve this in practice. This might be done by comparing the worldviews of cohorts of individuals (e.g. school children, university students, general public), ideally from a range of cultural backgrounds, before and after exposure to cosmic and evolutionary perspectives. I am not aware of any such studies, and I don't have the expertise to advise on appropriate methodologies for them, but I do think they would be worth performing.[12]

### Geopolitical Implications

The importance of developing a planetary perspective as a prerequisite for effectively tackling planetary-scale problems has long been recognized in the professional international relations community (e.g. Morgenthau, 1948; Herz, 1962; Ward, 1966). The potential role of big history in developing this perspective, with geopolitical implications, has recently been noted by Jo Leinen and Andreas Bummel (2018) in their book *A World Parliament: Governance and Democracy in the 21st Century* (p.361):

> Big history provides an account of the origin of all existence and of life on Earth on a strictly scientific basis. The cosmological worldview thus helps us on the path to an integral consciousness and creates an important frame of reference for planetary identity.

---

9   The phrase "Mankind is made of star stuff" is often attributed to Carl Sagan, but as far as I am aware Shapley was the first to use it.

10  Emphasis in the original.

11  Although the title of Goodenough's book suggests a theistic outlook, her actual perspective is one of 'religious natural-ism' which combines a naturalistic worldview with emotion-al and ethical perspectives normally associated with religion. As she argues (p. xiv), "the role of religion is to integrate the cosmology and the morality" of a culture. It seems important to recognize that if the 'Epic of Evolution' (aka big history) is perceived to be consistent with at least some religious worldviews that may aid its wider acceptance, although big history itself is better seen (in David Christian's phrase) as a secular 'origin story' anchored in scientific fact.

12  I am grateful to an anonymous reviewer for this suggestion.

The need for such a perspective is also developed in the *Planet Politics Manifesto* advanced by Anthony Burke and colleagues (2016). They argue that the existing, state-centric, political organisation of the world is "failing the reality of the planet", and seek to reorientate the study of international relations to answer the question "Can we match the planet with our politics?" They conclude that:

> Our fundamental image of the world must be revolutionised. Our existence is neither international nor global, but planetary. Our anthropocentric, state-centric, and capital-centric image of international relations and world politics is fundamentally wrong; it perpetuates the wrong reality, the wrong commitments and purposes, the wrong 'world-picture'.

Importantly, they stress that in order to make progress "we don't need more reports or policy debates. We need new practices, new ideas, stories and myths." By providing a common, scientifically robust, "origin story" (or, viewed another way, a "myth" describing humanity's place in the universe that is as true as modern science can make it), big history and related disciplines can help satisfy the last two of Burke et. al.'s prerequisites for progress, while in parallel stimulating interdisciplinary advances in the first two.

It is interesting to consider the potential longer-term political implications of a "planetary identity" engendered (in part) by big history. Fred Spier has drawn attention to the fact that academic history in its modern form emerged in the 19th century, largely to support the formation and consolidation of nation-states, and that this nationalistic imperative has led to the downplaying of integrated human, or universal, histories. This then leads him (Spier, 2015; p. 12) to make the following observation:

> the study of human history as a whole has only rarely been practiced up to the present. This remarkable situation may be linked to the fact that to do so would produce global identities, which are not directly associated with any presently viable state society.

This begs the question, already alluded to in the title of Leinen and Bummel's book quoted above, of whether the creation of "global identities" through the promulgation of big history and related perspectives could help in the development of global political institutions above the level of the nation-state. Both Wells and Shapley were convinced of this, and both devoted chapters of their books to making the case for world government[13]. Moreover, although authors like Wells and Shapley might easily be dismissed as overly idealistic and lacking in professional expertise in the field of international relations, essentially the same conclusion was reached by such leading 'realist' international relations scholars as Hans Morgenthau (1948) and John Herz (1962). Daniel Deudney (2018) has recently summarised Morgenthau's position as follows: "humanity thus faces a tragic impasse: it needs a world state for security, but lacks a sufficiently thick sense of common identity both to make it possible and to prevent it from being threatening." Morgenthau himself (1948, p. 419) appears to have viewed this as a challenge to be overcome:

---

13  Wells (1920) Chapter XLI: "The possible unification of the world into one community of knowledge and will"; Shapley (1963) Chapter 13: "The coming world state."

If the world state is unattainable in our world, yet indispensable for the survival of that world, it is necessary to create the conditions under which it will not be impossible from the outset to establish a world state.

Morgenthau saw the way forward through international diplomacy, but was clearly aware that developing a sense of common identity would be a prerequisite for success, just as "the community of the American people antedated the American state … a world community must antedate a world state (Morgenthau, 1948; p. 406).

This is not the place to reiterate all the arguments for or against the creation of a world government, or the various forms such a government might take. There is a large literature on this topic to which the interested reader can refer (e.g., Kant, 1795; Russell 1916; Laski, 1925; Reves, 1946; Toynbee, 1972; Kerr, 1990; Hamer, 1998; Wendt, 2003; Baratta, 2004; Yunker, 2007; Cabrera, 2011; Wendt, 2015; Leinen & Bummel, 2018; Hamer, this volume); a comprehensive and scholarly historical overview has been given by Heater (1996), and interested readers may wish to follow the contemporary on-line discussions at the *World Government Research Network*.[14] My own view (e.g. Crawford, 2015; esp. pp. 206-209) is that a federal world government, implementing the principle of subsidiarity[15] on a global scale, would be the most appropriate institutional response to tackling the many planetary-scale problems that human civilisation will face in the 21st century. That said, I find myself in agreement with Morgenthau and others that such geopolitical developments, while desirable, may be impractical until humanity develops a greater sense of its common identity, what Herz (1962, p. 317) termed a "planetary mind", Anderson (1991, p. 6) a sense of

"imagined community", and Ward (1966, p. 148) "a patriotism for the world itself".[16]

It seems to me that the temporal and evolutionary perspectives provided by big history, combined with the spatial ('cosmic') perspectives provided by the exploration of space (discussed below), will play a valuable, and perhaps essential, role in laying the foundations for a common human identity on which a future world government might be built (see also Crawford, 2018b).

## Space Exploration: Augmenting the Cosmic Perspective

Big history and astrobiology are both concerned with the *future* of humanity as well as the past, and, barring some unforeseen calamity, it seems likely that the exploration of space will be a part of this future. Certainly, if some of the more ambitious aspirations to make humanity a multi-planet species are realised, space exploration and development could become a very large part of the human (and post-human) future. Even if these aspirations are never realised, it seems likely that we will continue to explore our Solar System with robotic space probes, and probably also with astronauts. In this section I will therefore briefly examine the synergies, as I see them, between astrobiology, big history, and the exploration of space. Of course, space exploration is already an important component of astrobiology, because space probes are required to search for life on other planets, and discoveries made by space probes and space telescopes also inform big history. However, beyond these essentially practical synergies, I contend that important socio-political benefits will also result from an ambitious programme of space exploration, and that these will reinforce the societal benefits of big

---

14  http://wgresearch.org/ (accessed 17 December 2018).

15  I.e., that "a central authority should have a subsidiary function, performing only those tasks which cannot be performed effectively at a more immediate or local level" (OED, 2013).

16  Barbara Ward (aka Baroness Jackson)'s slim book *Spaceship Earth* (1966), based on her George P. Pegram lectures at Columbia University, contains much of interest to the present discussion. Of particular importance is her insistence on the need to build global institutions for planetary management.

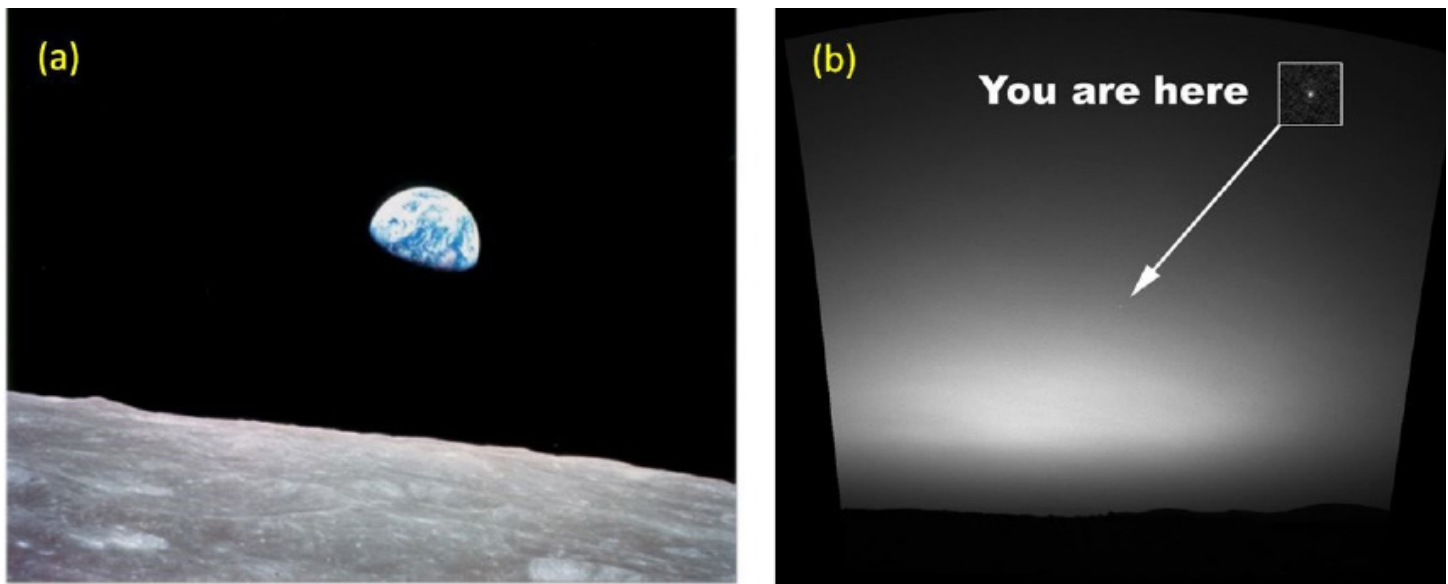

**Figure 2**. The cosmic perspective: (a) Earthrise over the lunar surface, photographed by the crew of Apollo 8 in December 1968. (b) The Earth photographed from the surface of Mars by the Mars Exploration Rover Spirit in March 2004. Such images powerfully reinforce a 'cosmic perspective' that can have a unifying influence on human affairs. Images courtesy of NASA.

history and astrobiology discussed above.

Most importantly, space exploration provides a *spatial* perspective on human affairs which complements the temporal and evolutionary perspectives of big history. Any society that is rigorously exploring the Solar System, can hardly fail to be aware that Earth is a very small planet when viewed in its cosmic setting (Figure 2). The social, cultural and psychological importance of this perspective has been noted by multiple authors (e.g. Clarke, 1946, 1951; Hoyle, 1950; Ward, 1966; Sagan, 1994; Poole, 2008; White, 2014). For example, even before any images of Earth from space had been obtained, the astronomer Fred Hoyle (1950, p. 9) wrote that:

> Once a photograph of the Earth, taken from the outside, is available, we shall, in an emotional sense, acquire an additional dimension … once let the sheer isolation of the Earth becomes plain to every man, whatever his nationality or creed, and a new idea as powerful as any in history will be let loose.

There is persuasive evidence that images of the Earth from space have raised environmental awareness, and thus contributed to popular movements for the reduction of pollution and the preservation of biodiversity (e.g., Zimmerman, 1998; Poole, 2008; Spier, this volume)[17]. Although it is sadly true that the cosmic perspective of "Spaceship Earth" (Ward, 1966; Fuller, 1969) hasn't yet triggered a sufficiently strong global response to solve these environmental problems, raising awareness of their planetary scale is nevertheless an important contribution of space exploration and a prerequisite for political action.

---

17 Zimmerman (1998, p. 275) reproduces an interesting diagram from Balzhiser (1990) which shows a dramatic growth in US environmental legislation in the late 1960s; proving a causal link to images of the Earth taken from space may not be possible, but the timing is suggestive. Fred Spier (this volume) draws attention to the differences in cultural impact of the original Apollo 8 'Earthrise' image (Fig. 2(a) above) in the United States and Europe; he argues that the immediate impact, especially outside of the US, may not have been as great as is often assumed, although its legacy has proved to be lasting and influential.

Similar observations can be made regarding the geopolitical implications of the cosmic perspective. Even before the space age, the science fiction author and space visionary Arthur C. Clarke (1946, p. 72) had noted that:

> It is not easy to see how the more extreme forms of nationalism can long survive when men begin to see the Earth in its true perspective as a single small globe among the stars.

Hoyle (1950, p. 9) echoed this sentiment a few years later, when he noted that this new perspective "must increasingly have the effect of exposing the futility of nationalistic strife." By the 1960s, when images of the Earth from space had been obtained, the implications were not lost on at least some professional diplomats. For example, Adlai Stevenson, then US Ambassador to the United Nations, expressed his view (Stevenson, 1965), that "we can never again be a squabbling band of nations before the awful majesty of outer space." This perspective is, understandably, much more visceral for people who have actually seen our planet from outside (White, 2014), and it is worth quoting one such observation here:

> You look down there and you can't imagine how many borders and boundaries you cross, again and again and again, and you don't even see them. There you are – hundreds of people in the Mid-East killing each other over some imaginary line that you're not even aware of …. And from where you see it the thing is a whole, and it's so beautiful. You wish you could take one in each hand, one from each side in the various conflicts, and say, 'Look. Look at it from this perspective….' (Schweickart, 1977).

As space exploration proceeds more people will be exposed to this perspective, both in person and vicariously, and the more it will diffuse through society. Such an enlargement of perspective can hardly fail to strengthen the sense of planetary identity inherent in big historical and astrobiological worldviews. Indeed, images of Earth from space, and especially personal experiences of this perspective, are likely to be even more effective in this regard because they prompt an instinctive, emotional, appreciation of 'one worldness' that the more intellectual perspectives provided by big history, astrobiology, and related academic disciplines cannot. We may hope that this perspective will gradually gnaw at the minds of political leaders (as it clearly did for Adlai Stevenson), and the minds of the wider public, until it leads to the emotional realisation that human activities affecting the planet as a whole need, and *ought*, to be organised collectively (see, e.g., Crawford, 2017). Only space exploration can provide this perspective, which has led Frank White (2014, p. 102) to argue that:

> It is time for the influence of space exploration on human consciousness to be seen as a legitimate justification for investing in it.

## Cultural Benefits of Space Exploration

In addition to providing a valuable, and uniquely compelling, spatial perspective on human existence, an ambitious future programme of space exploration will also result in a range of additional social and cultural benefits. Leaving aside the strictly scientific benefits, to which the whole history of space exploration can attest, I think we can also identify potential cultural benefits of space exploration under the broad headings of 'art', 'philosophy', and, albeit in the more distant future, 'diversity'. I have addressed these aspects in previous publications (e.g., Crawford, 1993, 2014), which I summarise here.

William McLaughlin (1993) considered the potential impact of space exploration on the fine arts and concluded that the influence is likely to be considerable. At one level it seems obvious that new space scenes, and novel space events and experiences, must inspire new works of space art. It is difficult to see how this could be otherwise. However, the potential long-term artistic impact of space exploration is likely to be more profound. The increasing dominance of the cosmic perspective on human thought is likely to change the whole paradigm of artistic expression. Not only will it be necessary to find new ways of portraying and communicating human (and human-derived) values in the face of a universe whose strangeness will likely become ever more apparent as exploration proceeds, but the human (and post-human) mind is itself likely to become increasingly 'cosmicized' (Finney, 1988) in a way that can hardly fail to be reflected in artistic and cultural evolution. Indeed, in the immediate aftermath of the Apollo missions to the Moon, the American literary scholar Joseph Campbell (1972, p. 233) clearly grasped this insight when he wrote:

> For although our voyage is to be outward, it is also to be inward, to the sources of all great acts, which are not out there, but in here, in us all, where the muses dwell.

And, further (p. 236) that:

> All the old bindings are broken. Cosmological centers now are any- and everywhere…all poetry now is archaic that fails to match the wonder of this view.

If anything, the stimulus that space exploration will provide for the philosophical disciplines may be even more profound. In Table 2, I summarise some philosophical issues that are likely to be stimulated as humanity (and post-humanity) moves out into the Solar System, and perhaps beyond. I have made a distinction between natural, moral and political philosophy, but we must also expect that the vast and mysterious universe in which we live very likely contains the seeds of entirely new fields of philosophical investigation waiting to be discovered.

In the longer term, one of the most important socio-cultural contributions of space exploration may be the opportunities it will provide for increasing human (and post-human) cultural diversity. In the nineteenth century, John Stuart Mill drew attention to the benefits of what he termed "different experiments of living" (Mill, 1859; p. 120), but such experiments are becoming increasingly difficult in a homogenizing world. Indeed, I have argued above that some of this homogenization, at least on a political level, is positively desirable if it helps breakdown tribal animosities on Earth, and that a common 'big historical' perspective could help facilitate this. Moreover, although federal political systems, such as a future federal world government, are well-suited to maintaining cultural diversity in the face of common high level political structures, it seems likely that cultural diversity on this planet is likely to continue to decrease.

Although clearly a long way in the future, it is possible that space exploration, and especially the colonisation of other planets by humans (and post-humans), will provide a solution to this dilemma. Interestingly, this possibility was recognized by the philosopher Olaf Stapledon (1948) a decade before the space age had even begun, when he expressed the view that:

> The goal for the solar system would seem to be that it should become an interplanetary community of very diverse worlds each inhabited by its appropriate race of intelligent beings, its characteristic "humanity"… Through the pooling of this wealth of experience, through this 'commonwealth of worlds' new levels of mental and spiritual development should become possible, levels at present quite inconceivable to man.[18]

That said, the colonisation of the Solar System will also create additional risks: we don't want to unite the Earth only to live in a politically anarchic Solar System where colossal energies would be

18  Much of Stapledon's thought is relevant to big historical and astrobiological perspectives, and I recommend especially his science fiction novel *Star Maker* (Stapledon, 1937). For a more detailed discussion of Stapledon's ideas in the context of space exploration, see Crawford (2012).

| Natural Philosophy | Moral and Ethical Philosophy | Political Philosophy |
|---|---|---|
| How secure is our basic physical understanding of the universe? | Extension of environmental ethics to other planets. | Consideration of the ownership of extraterrestrial resources |
| Can we define 'life' in a cosmic context? Is this even important? | What are the moral and ethical relationships between humanity and extraterrestrial life (should any be encountered)? | Consideration of appropriate forms of planetary and interplanetary governance. |
| If life can be defined, how common is it in the universe? What are the ultimate constraints on the origin of life and its distribution? | What are the ethical implications of spreading Earth-life through the Solar System and the Galaxy? | Consideration of political relationships with advanced extraterrestrial societies (if any); what limits would *biological* differences place on developing political institutions? |

**Table 2**: Some philosophical issues that are likely to arise as space exploration proceeds.

available for anyone (or anything) minded to use them destructively (e.g., Baxter and Crawford, 2015; Deudney, 2016, 2019) For this reason, care will have to be given to developing appropriate interplanetary political institutions (Crawford, 2015).

Thirty years ago, the American political philosopher Francis Fukuyama (1989, 1992) argued that our world is becoming politically and culturally homogenized, and that this may lead to political and cultural stagnation. Following Hegel (1832), Fukuyama famously (or, depending on your point of view, infamously) termed this perceived endpoint in human cultural evolution the 'End of History'. Although subsequent events have shown that this process is proceeding more slowly than Fukuyama perhaps envisaged, some of the trends he identified seem likely to continue. Although, as I have argued above, increasing *political* unification of humanity seems positively desirable, Fukuyama's concerns regarding cultural stagnation in a politically unifying world do need to be taken seriously. As he put it (Fukuyama, 1989, p. 18):

The end of history will be a very sad time. The struggle for recognition, the willingness to risk one's life for a purely abstract goal, the worldwide ideological struggle that called forth daring, courage, imagination, and idealism, will be replaced by economic calculation, the endless solving of technical problems, environmental concerns, and the satisfaction of sophisticated consumer demands. In the post-historical period, there will be neither art nor philosophy, just the perpetual caretaking of the museum of human history.

A decade before the dawn of the space age, the possibility that an ambitious programme of space exploration could help prevent just this kind of cultural and intellectual stagnation was recognized by Clarke (1946, p. 72) when he wrote:

Interplanetary travel is the only form of 'conquest and empire' now compatible with civilisation. Without it, the human mind, compelled to circle forever in its planetary goldfish bowl, must eventually stagnate.

Human expansion into the Solar System, and eventually beyond, will certainly present a vast new field of human activity, with literally infinite potential for discovery and intellectual stimulation on multiple levels.

As Dunér (2013, p. 13) has recently argued:

> Encounters with the unknown outer space will … change our thinking, conceptions, categories, belief systems, culture and meanings of things. What we have come to believe so far through science and human cognition will face anomalies. The old categories, systems, and beliefs will fall short when we try to understand these new unfamiliar things. Our thinking, science, and belief systems will then have to be revised.

However one views it, it seems certain that a future in which space exploration plays a significant role will provide a far richer range of cultural and intellectual stimuli than we could ever hope to experience if we never leave our home planet (e.g., Clarke, 1946, 1951; Sagan, 1994; Crawford, 2014). Sagan (1994, p. 285) perhaps expressed it as well as anyone:

> We're the kind of species that needs a frontier – for fundamental biological reasons. Every time humanity stretches itself and turns a new corner, it receives a jolt of productive vitality that can carry it for centuries.

In the long run, the exploration of space may help us avoid Fukuyama's 'End of History' by keeping history *open* while simultaneously helping to unite human cultures on Earth.

## Conclusions

The twin, and closely related, academic disciplines of big history and astrobiology have the potential to yield a wide range of social and intellectual benefits. Indeed, intellectual enrichment is already resulting from the interdisciplinary research agendas of both astrobiology and big history, which involve scholars from a wide range of sciences and the humanities working closely together. More importantly, both disciplines rely on, and naturally engender, cosmic and evolutionary perspectives which, I argue, ought to form part of the worldview of every educated person (see also Elise Bohan's paper in this volume).

If suitable methodologies could be conceived and implemented, it would be desirable to quantify the effects of exposure to these perspectives on individuals from a wide range of ages and cultural and educational backgrounds. Such data could then inform evidence-based proposals for reforming educational curricula to include big history and related mind-broadening perspectives.

By powerfully reinforcing the fact that all human beings, and all human societies, exist on the same small planet, and are related by a common evolutionary history, I have argued that cosmic and evolutionary perspectives strengthen intellectual and emotional arguments for the eventual political unification of humanity. My own view is that a federal world government would be an appropriate institutional framework for a united humanity, and that a world government of some kind may be necessary if serious global problems are to be properly managed. However, such a political outcome is only likely to become realistic if humanity develops a greater sense of its common identity, what Barbara Ward (1966, p. 148) called "a patriotism for the world itself." The perspectives provided by big history, astrobiology and space exploration can all help achieve this objective. That said, I also agree with Fukuyama (1989) that a politically homogenised world may lack sufficient sources of intellectual stimuli to maintain a vibrant culture, and I have argued that an ambitious programme of space exploration would help in this respect. Needless-to-say, the exploration of space will also yield new knowledge about the universe, informing both the science of astrobiology and the ever-evolving big historical worldview.

## Acknowledgements

This paper was presented at a meeting on the theme 'Expanding Worldviews: Astrobiology, Big History, and the Social and Intellectual Benefits of the Cosmic Perspective' that was held on 19 July 2018 under the auspices of the Humanities Research Centre (HRC) at the Australian National University, and was largely written while I held a Visiting Fellowship at the HRC. I thank the HRC, and especially Professor Will Christie and Ms Penny Brew, for hosting the meeting and for their hospitality during my stay as a Visiting Fellow. I also thank an anonymous reviewer for comments on the manuscript that have improved it.